\documentstyle{mn}
\newif\ifAMStwofonts

\input epsf

  \ifCUPmtlplainloaded \else
  \fi
  \ifAMStwofonts
    \ifCUPmtlplainloaded \else
      \NewSymbolFont{upmath} {eurm10}
      \NewSymbolFont{AMSa} {msam10}
      \NewMathSymbol{\upi}     {0}{upmath}{19}
      \NewMathSymbol{\umu}     {0}{upmath}{16}
      \NewMathSymbol{\upartial}{0}{upmath}{40}
      \NewMathSymbol{\leqslant}{3}{AMSa}{36}
      \NewMathSymbol{\geqslant}{3}{AMSa}{3E}

      \let\leq=\leqslant \let\le=\leqslant
       \let\ge=\geqslant
    \fi
  \fi

\ifnfssone
  \newmathalphabet{\mathit}
  \addtoversion{normal}{\mathit}{cmr}{m}{it}
  \addtoversion{bold}{\mathit}{cmr}{bx}{it}
  \newmathalphabet{\mathbfit} 
  \addtoversion{normal}{\mathbfit}{cmr}{bx}{it}
  \addtoversion{bold}{\mathbfit}{cmr}{bx}{it}
  \newmathalphabet{\mathbfss} 
  \addtoversion{normal}{\mathbfss}{cmss}{bx}{n}
  \addtoversion{bold}{\mathbfss}{cmss}{bx}{n}
  \ifAMStwofonts
    \ifCUPmtlplainloaded \else
      \UseAMStwoboldmath
      \makeatletter
      \new@mathgroup\upmath@group
      \define@mathgroup\mv@normal\upmath@group{eur}{m}{n}
      \define@mathgroup\mv@bold\upmath@group{eur}{b}{n}
      \edef\UPM{\hexnumber\upmath@group}
      \new@mathgroup\amsa@group
      \define@mathgroup\mv@normal\amsa@group{msa}{m}{n}
      \define@mathgroup\mv@bold\amsa@group{msa}{m}{n}
      \edef\AMSa{\hexnumber\amsa@group}
      \makeatother
      \mathchardef\upi="0\UPM19
      \mathchardef\umu="0\UPM16
      \mathchardef\upartial="0\UPM40
      \mathchardef\leqslant="3\AMSa36
      \mathchardef\geqslant="3\AMSa3E

      \let\leq=\leqslant \let\le=\leqslant
       \let\ge=\geqslant
    \fi
  \fi
\fi 

\ifnfsstwo
  \DeclareMathAlphabet{\mathbfit}{OT1}{cmr}{bx}{it}
  \SetMathAlphabet\mathbfit{bold}{OT1}{cmr}{bx}{it}
  \DeclareMathAlphabet{\mathbfss}{OT1}{cmss}{bx}{n}
  \SetMathAlphabet\mathbfss{bold}{OT1}{cmss}{bx}{n}
  \ifAMStwofonts
    \ifCUPmtlplainloaded \else
      \DeclareSymbolFont{UPM}{U}{eur}{m}{n}
      \SetSymbolFont{UPM}{bold}{U}{eur}{b}{n}
      \DeclareSymbolFont{AMSa}{U}{msa}{m}{n}
      \DeclareMathSymbol{\upi}{0}{UPM}{"19}
      \DeclareMathSymbol{\umu}{0}{UPM}{"16}                             \DeclareMathSymbol{\upartial}{0}{UPM}{"40}
      \DeclareMathSymbol{\leqslant}{3}{AMSa}{"36}
      \DeclareMathSymbol{\geqslant}{3}{AMSa}{"3E}

      \let\leq=\leqslant \let\le=\leqslant
       \let\ge=\geqslant
    \fi
  \fi
\fi 

\ifCUPmtlplainloaded \else
  \ifAMStwofonts \else 
    \def\upi{\pi}
    \def\umu{\mu}
    \def\upartial{\partial}
  \fi
\fi

\begin{document}
\title{Asymptotic Giant Branch Stars as Astroparticle Laboratories}

\author [Inma Dom\'\i nguez el al.]{Inma Dom\'{\i}nguez,$^1$ Oscar Straniero,$^2$ and Jordi Isern $^3$\\
$^1$Depto. de F\'{\i}sica Te\'orica y del Cosmos, Universidad de Granada,
18071 Granada, Spain (inma@ugr.es)\\
$^2$Osservatorio Astronomico di Collurania,
64100 Teramo, Italy (straniero@astrte.te.astro.it)\\
$^3$Institut d'Estudis Espacials de Catalunya, IEEC/CSIC, 
 Barcelona, Spain (isern@ieec.fcr.es)}
\maketitle
\begin{abstract}

We show that the inclusion of axion emission during stellar 
  evolution introduces important changes into the evolutionary behaviour of AGB
 stars. The mass of the resulting C/O white dwarf is much lower than the equivalent obtained from standard evolution. This implies a 
 deficit in luminous AGB stars and in massive WDs. Moreover   
 the total mass processed in the nuclear burning shells that
is dredged-up to the surface ($3^{rd} D_{up}$)  increases when axion emission is
included, modifying  the chemical composition of the photosphere.  
We conclude that the AGB is a promising phase to put constraints on particle physics  
\end{abstract}
\begin{keywords}Axions--Stellar Evolution--AGB--WD.
\end{keywords}

\section {Introduction}

It is well known that the standard theory does not  predict 
 several properties of  
the elementary particles (such as mass and  magnetic moment). Furthermore,  
 the different  non-standard
 theories leave open the possibility 
that unknown (exotic) particles could still exist. In the recent past, 
several attempts have been made  to understand how these particles 
could modify  stellar evolution and, in turn, to use these modifications 
 to  constrain particle theory (see Raffelt 1996 for a recent review).
  The general procedure consists in the comparison of the observed properties of selected stars (or of a cluster of stars) with the prediction of theoretical stellar models obtained under different assumptions about the microphysics input.

Axions and WIMPS (Weak Interactive Massive Particles) are the most promising candidates for non baryonic dark matter. Non accelerator tests for 
 axions  in the acceptable mass range are up to date available.
 In this situation, stars in general and especially the best-known one,  
  the Sun, are being widely used to test particle physics. 
 Several experiments to detect  solar and  galactic axions have been performed (Sikivie 1983; Lazarus et al. 
 1992). Experiments with better  sensitivity  are currently underway 
  (van Bibber et al.  1994; Hagmann et al. 1996 and 1998 (U.S. Axion Search); 
  Matsuki et al. 1996 (Kyoto experiment CARRACK); Moriyama 1998, Moriyama et al. 1998; Avignone III et al. 1998 and Gattone et al. (SOLAX collaboration)) and there is a  proposal  to use 
 the Large Hadron Collider at CERN (Zioutas et al. 1998) to detect solar axions.    
For the moment,  results have proved to be  negative and therefore the existence of axions is 
 an attractive 
 but speculative hypothesis. 

The main purpose of the present paper is to show that  Asymptotic Giant Branch (AGB) stars can also be used  
successfully as astroparticle physics laboratories. Since they are very bright,  their photometric and
spectroscopic properties are well known; for example, the   AGB luminosity function
 can be used to test 
the efficiency of any phenomenon which is related with the production,
removal  or transport of energy. Furthermore, the possibility of observing 
 directly the ongoing nucleosynthesis  provides a rare opportunity
to obtain information about the internal physical conditions. 
We recall that AGB stars are formed by a dense
and degenerate core made up of carbon and oxygen (the CO core) surrounded by two interacting burning shells.
The typical densities 
and temperatures of the CO core and of  the He and H-rich layers  
make these stars a suitable environment to check the
reliability of the theory of nuclear and particle physics.  

In order to illustrate the possible use of  AGB stars in the framework of astroparticle research,
we will address the question of the existence of axions (Peccei
and Quinn 1977a and 1977b). There are two types of axion models, the hadronic
model or KSVZ (Kim 1979, Shifman et al. 1980) and the  DFSZ  
(Zhitniskii 1980, Dine et al.  1981) model. In the first model, the axion couples to
hadrons and photons and in the second one also to charged leptons.
 The axion  mass is 
 $m_{\rm ax}=0.62\,{\rm eV}
(10^7 \,{\rm GeV}/f_{\rm  a})$, where $f_{\rm  a}$ is the Peccei--Quinn
scale and the axion coupling to matter is proportional to $f_{\rm  a}^{-1}$. 
In principle, the model does not put any  constraint  on the value of 
 $f_{\rm  a}$, so the limits to the $m_{\rm ax}$ must be  obtained from 
 experimental arguments. 
 
    Within the accepted mass range of DFSZ axions, 
  10$^{-5}$ eV $\le m_{\rm  ax} \le$ 0.01 eV (see Raffelt, 1998), axions produced in stellar
interiors can freely escape from the star and remove energy, as do 
 neutrinos.
The possibility of using stellar interiors to constraint the axion mass 
  was 
 early recognised (Fukugita et al. 1982).  
 The axion energy loss rates  depend on the axion coupling strenght  to 
 electrons, photons and nucleons and the corresponding coupling constants   depend on the
 axion 
 mass and other model dependent constants. In this way 
 axion mass limits are obtained.

Axions may couple to electrons (Kim 1987; Cheng 1988; Raffelt 1990; 
Turner 1990 and Kolb and Turner 1990); the  axionic fine
structure constant is taken as 
 $\alpha = g^2/4\pi$, 
 g is a dimensionless coupling constant that in the DFSZ 
 models is $g= 2.83 \times 10^{-8} m_a/cos^2\beta$, 
  $m_a$ is the axion mass in meV (10$^{-3}$ eV) and $cos^2\beta$ 
is a model-dependent parameter that is usually set equal to 1. 

In  models in which axions couple to electrons, the two most interesting types of axion interaction that can  occur in stellar interiors are photoaxions  ($\gamma + e \rightarrow e+a$), which is a particular branch 
of the Compton scattering, and
bremsstrahlung axions ($e + [Z,A] \rightarrow [Z,A] + e +a$)(see Raffelt, 1990).
Figure 1 illustrates  the axion
energy loss rates for these two kinds of interactions, for two chemical 
 compositions. Compton emission strongly increases 
with temperature (approximately as T$^6$) and is almost independent of the density. It is,  however, inhibited by electron degeneracy. Thus, Compton emission is
important in non-degenerate He-burning regions, when the temperature and, in turn, the photon density
are high enough. Bremsstrahlung requires greater densities and  is not damped at  large  electron degeneracies.
Thus, it is dominant in degenerate He and CO cores.

As yet, few works have  included  axion interactions in stellar
model computations (see  Raffelt 1996 and references therein).  
Isern, Hernanz \& Garc\'{\i}a--Berro (1992) use the rate of change of the pulsational period of
the white dwarfs during the cooling sequence to constrain the axion mass. Such a rate 
depends on the cooling rate, namely: $dt/d \ln P \propto dt/d \ln T$, where $P$
is the period of pulsation and T is the temperature of the white dwarf. If the
white dwarf has not entered  the crystallizing region of the cooling sequence, the discrepancy
between the rate of change of the period due to photons alone, which can be
obtained from models, and the observed one, which can be assumed to be caused
by photons and axions, is
$$
\frac{L_{\rm phot}+ L_{\rm ax}}{L_{\rm phot}} = \frac{\dot P_{\rm obs}}
{\dot P_{\rm mod}}
$$
\noindent
In such a way, Isern, Hernanz \& Garc\'{\i}a--Berro (1992 and 1993) found,  
 assuming $cos^2\beta$ = 1, that $m_{\rm ax} \la$ 8.7, 15 and 12 meV  by using the seismological  data of
 the white dwarfs G117--B15A, L19--2 and R548, respectively.

Other constraints can be obtained from   the possible influence of axions on 
the evolution of low mass red giants. These stars develop a degenerate He core in which axions,
if they exist, could produce  a significant energy loss. Dearborn, Schramm and
Steigman (1986) found that the helium ignition
would be suppressed for values of $m_{\rm ax}$ over a certain limit. 
 Unfortunatelly they overestimated the energy loss rate and later on, 
 Raffelt and Dearborn (1987) found that, with the current limits,  
   He ignition would never be supressed. These authors derived the most stringent  limit at that time, for the axion-photon coupling,  from the duration of the helium burning lifetime,  $m_{\rm ax}\la$  0.7 eV (KSVZ model). Raffelt and Weiss (1995) subsequently    
reexamined the
problem, analyzing the effects of the  axion-electron  coupling  in  
observable quantities, such as the luminosity of the RGB  (red giant branch) tip. They  concluded 
that $m_{\rm ax}$ is lower than 9 meV, assuming $cos^2\beta$=1 (DFSZ model).
Another limit was  obtained for the coupling strenght to photons  by 
 Raffelt (1996) by  
 comparing the number of    
  HB (horizontal branch)  with the number of 
RGB stars of galactic Globular Clusters, giving $m_{\rm ax}\la$  0.4 eV (DFSZ 
model). The most stringent limit come from the SN 1987A, constraints on the axion-nucleon coupling (Janka et al. 1996; Keil et al. 1997) give, taken into 
 account the overall uncertainties, $m_{\rm ax}\la$  0.01  eV for both, KSVZ and DFSZ models (Raffelt, 1998).  

If axions exist, this would be of major consequence on the evolution of AGB stars. In fact, 
in both the He shell and in the CO core the physical conditions for an important emission of axions are met.
Inside the core there are no nuclear reactions and the gain of
gravitational energy would be  balanced by neutrino and axion energy losses. The
He-burning shell reaches temperatures above  $3\times10^8$ K, much higher than those 
experienced during the central
He-burning, and this could induce a huge production of photoaxions.
Therefore axions could alter the stellar 
lifetime and change the final mass of the CO core.

In the following sections we will present various models of low and intermediate mass stars in the range of masses
$0.8 \le M/M_{\odot} \le 9$, with and without axion interactions. 
In particular we assume three different values
of the leading parameter 
$m_{\rm ax}$ (the axion mass), namely: 0, 8.5 and 20 meV (in the following Case 0, 1 and 2 respectively) and set always $cos^2\beta$=1.
Note that the larger value is roughly  
 double  the maximum value proposed by Raffelt (1998) yet  still
marginally compatible with the properties of white dwarfs. We show that the inclusion of axions would imply a significant
modification of the AGB characteristics. We do not address the question of
 setting an astrophysical 
upper limit of the axion mass here, since this requires a profound comparison 
 between  observations 
of galactic and Magellanic Clouds AGB stars
 and will be done in a forthcoming paper.

\begin{figure}
\epsfxsize=8 cm
\epsfbox{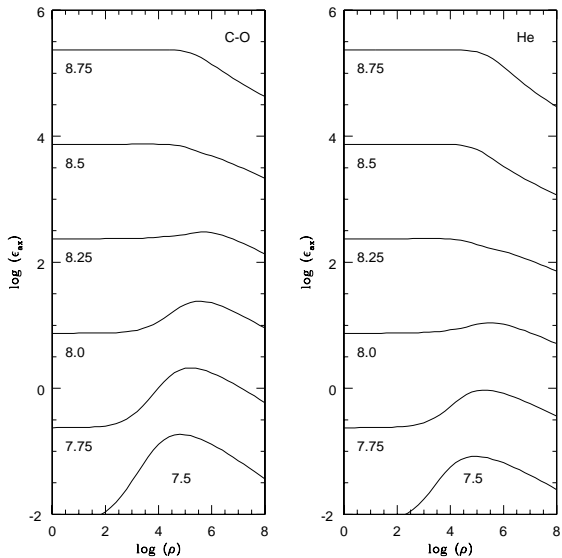}
\caption{Axion energy loss rate (erg/g/s) as a function of the density 
 (g/cm$^3$) for different temperatures (log(T) are indicated) and 2
 compositions, pure He and a mixture of  C and O (50\% each). 
The mass of the axion is assumed to be 8.5 meV.} 
\end{figure}

\section{The models}

We have followed the evolution of a set of low and intermediate mass stars 
(M= 0.8, 1.5, 3, 5, 7, 8 and 9 M$_\odot$) with solar metallicities, Z=0.02
and Y=0.28. This mass interval
includes stars that ignite He in degenerate conditions (0.8 and 1.5
M$_\odot$), stars that ignite He in non--degenerate conditions but form a
degenerate CO core (3--7 M$_\odot$) and go through the entire AGB phase, and
stars that ignite C in partially degenerate conditions (8 and 9 M$_\odot$).
The evolutionary sequences  are started at the ZAMS phase. 
For the two smaller masses (0.8 and 1.5 $M_{\odot}$)
we stopped the sequence at the onset of the He flash. All the other sequences   were terminated 
at the beginning of the thermal pulse phase or at C ignition. Due to the
large amount of computer time required to compute thermally pulsing models, 
only the 5 $M_{\odot}$ sequences (with and without axions) were  followed throughout this phase.

All the evolutionary sequences of stars were computed with the FRANEC code
(Frascati RAphson Newton Evolutionary Code), as described by Chieffi and
Straniero (1989). The most recent updates of the input physics were described 
by Straniero, Chieffi and Limongi (1997). When included, the axion energy loss rate have been computed 
according to Raffelt and Weiss (1995) in the case of Bremsstrahlung in degenerate conditions and Compton and according to Raffelt (1990) in the case of Bremsstrahlung in non degenerate conditions. 

 Finally, for the 5 M$_\odot$ star, we considered  the case in which only the CO core
experiences axion energy losses, in order to distinguish the relative effects on the core
and on the He shell and between the AGB and HB phases.

\section{Approaching the AGB}

Table 1 summarizes the results of the computed evolutionary
sequences.  From column 1 to 10 we report: the total mass (in solar units), the axion mass (in meV),
 the surface He mass
fraction after the first  dredge-up, the tip luminosity of the first RGB (red giant
branch), the mass of the He core at the beginning of the He-burning (in solar units), the central He
burning life time, the He core mass (in solar units) at the end of central He burning,
the E-AGB (early asymptotic giant branch) life time, the surface He mass fraction
after the second dredge-up and  the CO core mass (in solar units) at the end of the E-AGB.

\setcounter{table}{0}
\begin{table*}
\begin{minipage}{160mm}
\caption{Properties of the models during the HB and E-AGB phases}
\begin{tabular}{@{}cccccccccc}
    & & & & & & & & & \\
M$_T$ &
M$_{ax}$ &
He$_{s}^1$ &
log L$_{tip}$ &  M$_{He}^{1}$ &
$\Delta$t$_{He}$ &
M$_{He}^2$ &
$\Delta$t$_{E-AGB}$ &
 He$_{s}^2$ &
M$_{CO}$ \\
    & & & & & & & & & \\
    & 0.0  & 0.298 & 3.451 & 0.479 & ...  & ... & ... 
    & ...  & ... \\
 0.8   &  8.5 & 0.298 & 3.559 & 0.501 & ... & ...  &...  & ...
    & ...   \\
    & 20.  & 0.299 & 3.750 & 0.544 & ... & ... & ... & ...
    & ...   \\
    & & & & & & & & & \\
    & 0.0  & 0.294 & 3.445 & 0.477 & ... & ... & ... & ...
    &  ...  \\
 1.5   &  8.5 & 0.294 & 3.561 & 0.500 & ... & ... & ... & ...
    & ...   \\
    & 20.  & 0.295 & 3.765 & 0.547 & ... & ... & ... & ...
    &  ...  \\
    & & & & & & & & & \\
    &  0.0 & 0.296 & 2.560 & 0.378 & 141 &  0.545 & 
  9.2 & 0.296  & 0.549 \\
 3.0   &  8.5  & 0.296 & 2.587 & 0.378 & 127 &  0.530 &
  6.6 & 0.296  &  0.524  \\
    &  20.  & 0.298 & 2.707 & 0.378 & 85.6 &  0.479 &
  5.0 & 0.298 &  0.459 \\
    & & & & & & & & & \\
   &  0.0  &  0.296 & 3.186 & 0.654 &  20.8 &  1.024 &
  1.20 & 0.324 &  0.856 \\
 5.0  &  8.5  & 0.296 & 3.187 & 0.651 &  20.0 & 1.014 &
  0.75 &  0.340 &  0.744 \\
   &  20.  & 0.297 & 3.192 & 0.648 &  16.6 &  0.968 &
  0.51 & 0.347 &  0.645 \\
    & & & & & & & & & \\
   &  0.0  & 0.296 & 3.685 & 1.001 & 7.3 &  1.591 &
  0.39 & 0.366 & 1.005 \\
 7.0  &  8.5  & 0.298 & 3.686 & 1.001 & 7.05 &  1.579 &
  0.240 & 0.377 & 0.881 \\
   &  20.  &  0.299 & 3.688 & 1.001 & 6.2 &  1.537 &
  0.115 & 0.381 & 0.787 \\
    & & & & & & & & & \\
   &  0.0  & 0.300 & 3.881 & 1.201 & 5.1 &  1.882 &
 C-ignition  & ... & ... \\
 8.0  &  8.5  &  0.300 & 3.882 & 1.201 & 5.1 &  1.889 &
  C-ignition & ... & ... \\
   &  20.  &  0.301 & 3.883 & 1.201 & 4.64 &  1.585 & C-ignition 
    & ... & ... \\
    & & & & & & & & & \\
   &  0.0  & 0.301 & 4.052 & 1.422 & 4.1 &  2.217 &
  C-ignition & ... & ... \\
 9.0  &  8.5  &  0.301 & 4.053 & 1.422 & 4.1 & 2.209 &
  C-ignition & ...  & ... \\
   &  20.  &  0.302 & 4.055 & 1.422 & 3.9 & 2.165 &
  C-ignition & ... & ... \\
    & & & & & & & & & \\
    & & & & & & & & & \\
\end{tabular}
\end{minipage}
\end{table*}

During the central H-burning and up to the base of the RGB,  
the energy loss due to axions is negligible. 
However, when the temperature in the degenerate He core of a low mass star approaches $10^8$ K,
the contribution of the axion interactions to the energy balance becomes significant. 
In spite of the different initial chemical composition, our results for the 0.8 $M_{\odot}$ star are in good
agreement with those of Raffelt and Weiss (1995).
This is no  surprise since our axion rates are mainly taken from them.
For instance, they obtained an increment of the mass
of the He core at the onset of the He--flash of
0.022 M$_\odot$ and 0.056 M$_\odot$
for m$_{ax}$ = 8.9 and 17.9 meV respectively, while we obtain
0.021M$_\odot$ and 0.065 for axion masses of 8.5 and 20 meV respectively.

Note  that some widely used expressions like {\it delay to higher
densities} or {\it He ignition is delayed} might be misleading. The 
He ignition does indeed occur when the central density is higher,
but the density at the off centre ignition point in our calculations 
is lower (namely 30--40 $\%$ less in Case 2 than in Case 0). This point is situated further
from the centre in those models that include axions, since the maximum temperature
moves outwards as a consequence of the strong axion cooling in the
central region. For example, in the case of a 1.5M$_\odot$ star, the ignition
occurs at 0.14M$_\odot$, 0.25M$_\odot$
and  0.40M$_\odot$ in Cases 0, 1 and 2 respectively.
Similar differences  are obtained in the 0.8M$_\odot$.
Furthermore, He ignition is  not {\it delayed} in time, but {\it anticipated}.
The reason is that axion emission accelerates the contraction of the He core.
For instance, concerning the 0.8M$_\odot$, the RGB evolutionary time is about   
150 Myr ($\sim 6 \%$) lower in  Case 2 than in Case 0, while that of  1.5M$_\odot$  is 
68 Myr ($\sim 10 \%$) lower.
The luminosity at the 
RGB tip (column 4 ) is also increased (a factor $\sim$ 2 in Case 2) by axions 
in those stars that ignite He in a degenerate core.

The evolution of the more massive models (M $\ge 3 M_{\odot}$) is not affected by axions until 
the central He-burning is well established. 
As already found in low mass stars by Raffelt and Dearborn (1987), since the nuclear burning
must supply the energy lost by axions, the temperature is higher  and  He consumption is faster.
The reduction of the total evolutionary time of the He-burning phase 
(column 6) 
 varies from nearly 40$\%$ for the 3M$_\odot$ to 6$\%$ for the 9M$_\odot$ (for the higher value of the
axion mass, Case 2). In the models in which the smaller axion mass value was used, the reduction
is obviously less: 11$\%$ for the 3M$_\odot$ and negligible  for the
9M$_\odot$. Owing to the contraction of the evolutionary time, the H-burning shell has less
time to advance in mass and so the final He core mass is reduced (column 7 in Table 1). 
As a consequence of the higher  temperature in the He-burning core, the 
3$\alpha$ reaction is favoured with respect to the concurrent $\alpha$--burning of
$^{12}$C. Thus at the end of the He-burning, the abundance of carbon is greater. For instance,
in the 5M$_\odot$ model,
the central C abundance changes from 0.224 in Case 0 to 0.322 in Case 2.
On the whole, the larger the stellar mass the smaller the effect of  axion emission.

\section{The AGB}

The Asymptotic Giant Branch is characterized by two different phases, namely  the Early AGB (E-AGB)
and the thermal pulse phase (TP-AGB). During the E-AGB, the H and the He-burning shells are simultaneously
active. However  fuel consumption is faster in the inner He-burning shell 
 and so  the mass separation 
between the two burning regions becomes progressively smaller. The advancing  
He-burning induces an expansion and a cooling of the more external layers. If the stellar mass is large 
enough, the H-burning shell is finally quenched. In such a case, the convective envelope can penetrate 
the H/He discontinuity, bringing to the surface the products of the H-burning (basically N and He). 
This is called the second dredge-up and marks the end of the E-AGB and the beginning of the 
thermal pulses. During this second part of the AGB phase, the H-burning shell is reignited
while the He-burning one is quenched. Once a suitable amount of He is accumulated by the H-burning, 
a strong He-burning starts again, expands the external layer and again quenches the H burning. 
In a very short time
the He consumption has progressed such that the He shell again extinguished, the contraction takes place  
and H is reignited. This sequence repeats 
 until the mass loss removes the envelope and the star leaves the AGB. 
Note that during the pulse the region between the two shells becomes convectively
unstable and fully mixed. After the thermal pulse, when quiescent He-burning occurs and the H-burning shell is still extinguished, a new convective envelope penetration at the H/He discontinuity 
brings to the surface the products of the He burning (essentially carbon) and that of the neutron capture
nucleosynthesis (s-process, see e.g. Straniero et al. 1996, Gallino et al. 1998).
This is the so-called third dredge-up. 
The following sections  describe the modifications to this scenario 
 induced by axion emission.

\subsection{The early-AGB and the second dredge-up}

The most important effect of  axion emission during the E-AGB is the faster
consumption of fuel within the He-shell. This causes the
H-shell to be extinguished more quickly and the second dredge-up to occur earlier, resulting in a markedly smaller CO core mass. This behaviour is shown in Figure 2.

\begin{figure}
\epsfxsize=8cm
\epsfbox{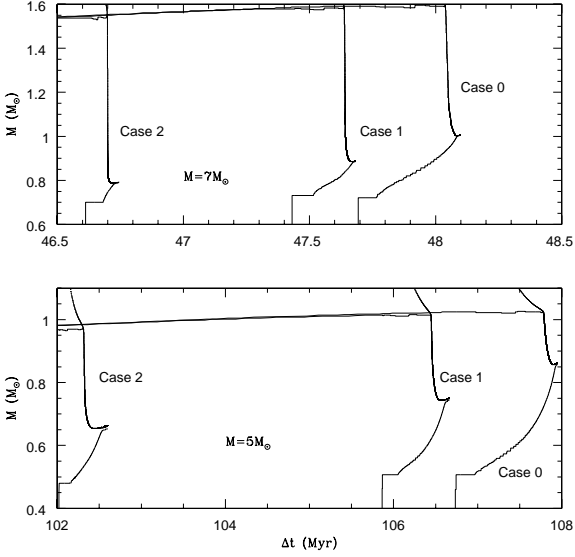}
\caption{Evolution with time of the C-O/He discontinuity,
 He/H discontinuity  and
the inner border (in mass coordinates) of the convective envelope for
    Case 0: m$_{ax}$=0.0, Case 1: 8.5 and Case 2: 20 meV.
Upper panel:  7M$_\odot$. Lower panel: 5M$_\odot$.}
\end{figure}

The reduction of the duration of the E-AGB phase (column 8, Table 1) is in
the range of  30--40$\%$, for Case 1; and as much  as 50--70$\%$ for Case 2.
The CO core mass at the end of the E-AGB is reported in column 10.
Note the significant differences with respect to the standard case. 
For the 7M$_\odot$ model, the CO core mass obtained for the highest axion mass is even smaller than that
obtained for  the standard 5M$_\odot$.
This reduction is not equal for all the masses, varying 
from  5$\%$ to 15$\%$ in Case 1, and from 15$\%$ to 28$\%$ in Case 2 for
stars in the mass range  3M$_\odot$ to 7M$_\odot$.

As a consequence of the deeper  penetration of the convective envelope,
the surface He abundance  after the 2$^{nd}$ dredge-up is greater (see column 9 in Table 1). 
For example, for the 5M$_\odot$ stars, the H/He discontinuity moves
inward  0.17M$_\odot$ in Case 0, and
0.30M$_\odot$  in Case 2.

As expected, the temperature within the CO core is lowered by
the larger energy loss.
Figure 3 clearly illustrates such an occurrence.

In order to elucidate how far these changes of the E-AGB phase are due to 
the previously altered  evolution (central He-burning) and/or to  
axion emission in the He shell, we have
limited their inclusion to the CO core. 
The result is that the
duration of the E-AGB is similar to  the one obtained in
the case of the full axion inclusion. 
This result is important, since it  clearly shows the strong
influence of the physical conditions in the degenerate core on the evolution
of the active shells. Furthermore, it means that the effects found during the E-AGB are
almost independent of the axion energy losses during the previous phases.

\begin{figure}
\epsfxsize=8cm
\epsfbox{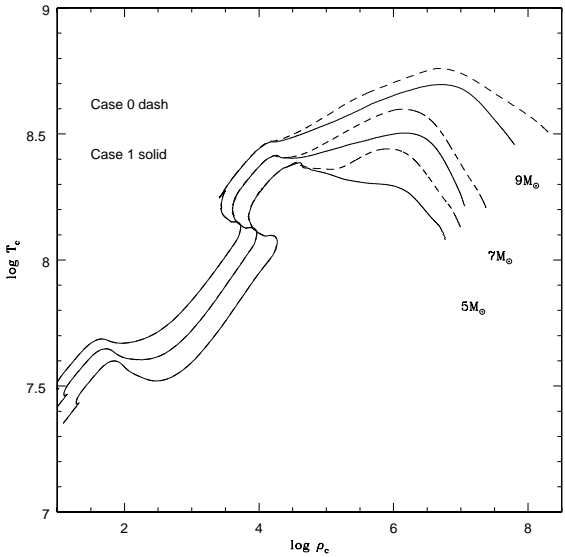}
\caption{Central temperature as a  function of the central density for the masses
9M$_\odot$ , 7M$_\odot$ and 5M$_\odot$.
Case 0: m$_{ax}$=0 meV; 
Case 1: m$_{ax}$=8.5 meV.}
\end{figure}

\subsection{The thermal pulse phase}

During the TP phase, axion emission in the core and in the He-burning shell becomes 
more important. In Table 2 we report some properties of the three sequences computed for the
5 $M_{\odot}$, namely the He core mass at the beginning of the TP (column 1), the duration of the interpulse 
period (column 2), the increment of the He core mass during the interpulse (column 3), the penetration 
(in solar masses) of the convective envelope at the time of the $3^{rd}$ dredge-up (column 4) and
the  $\lambda$ parameter, i.e. the ratio between the quantities in columns 3 and 4 (column 5).
In figure 4 we show the evolution of the surface luminosity. The evolution of the location of 
the external boundary of the CO core, of the He core and that of the internal boundary of
the convective envelope are shown in figure 5. 

The most striking consequence of  axion emission is the extension of the evolutionary time. 
The duration of a typical interpulse period
increases by a factor of $\sim$2.5 in Case 1 and by a factor of $\sim$6 in Case 2, both with respect to the standard case.
This is due to the combined effects of the lower core mass (see the previous subsection) 
and the axion cooling of the He shell.
The luminosity is about 1.5 times greater in Case 0 than in Case 2. However, 
note that all three sequences asimptotically follows the classical core mass/luminosity relation
(Pacinzsky 1977; Iben and Renzini, 1983). This is shown in figure 6 for Case 
 0 and 1. The temperature at the bottom of the 
convective envelope (figure 7) is lower in models which include axion emission. Thus the occurrence  
of hot bottom burning in the more advanced AGB evolution
and the consequent deviation from the core mass/luminosity relation
 (Bl\"ocker and Sch\"onberner 1991) should be reduced
by this axion cooling.

\begin{figure}
\epsfxsize=8cm
\epsfbox{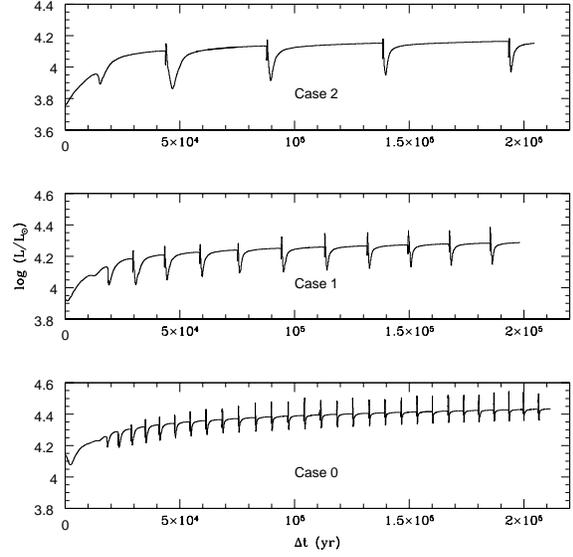}
\caption{Evolution with time of the surface luminosity during the TP-AGB
phase for the 5M$_\odot$ star. Lower panel: m$_{ax}$=0.0 (Case 0);
 Middle panel: m$_{ax}$=8.5 meV (Case 1); Upper panel:  m$_{ax}$=20 meV (Case 2).}
\end{figure}
\begin{figure}
\epsfxsize=8cm
\epsfbox{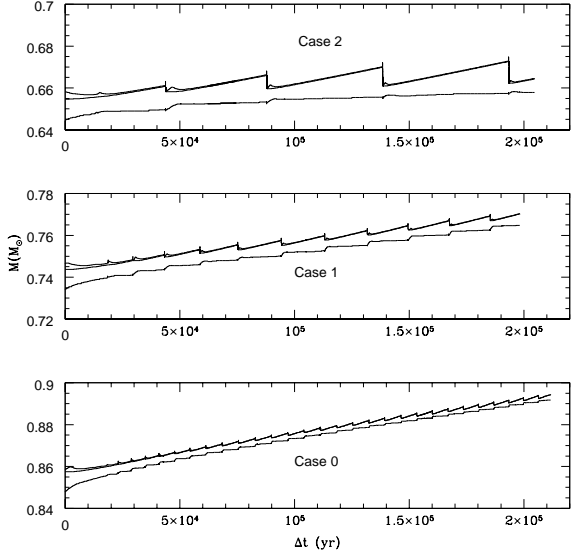}
\caption{Evolution with time of the C-O/He discontinuity, H/He
discontinuity and the inner border (in mass coordinates) of the convective envelope  at the TP-AGB phase, for the 5M$_\odot$ star.
Lower panel: m$_{ax}$=0.0 (Case 0);
 Middle panel: m$_{ax}$=8.5 meV (Case 1); Upper panel:
  m$_{ax}$=20 meV (Case 2).}
\end{figure}
\begin{figure}
\epsfxsize=8cm
\epsfbox{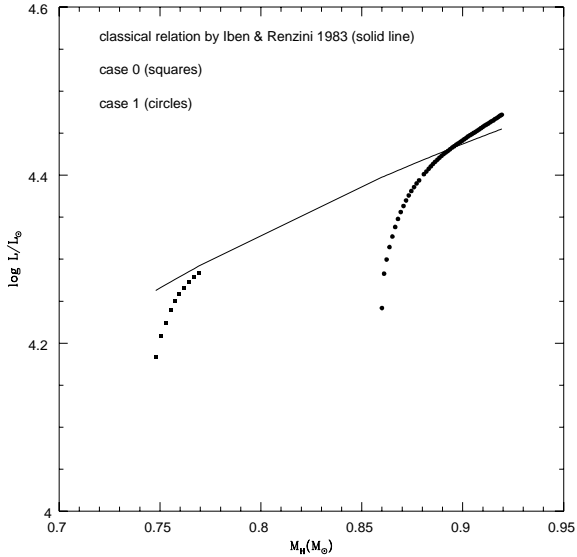}
\caption{Core Mass-Luminosity Relation for AGB stars. Continous 
line (IR83):  computed following   
Iben-Renzini (1983);  
Case 0:  the 5M$_\odot$ star with m$_{ax}$=0.0  and 
Case 1: same mass  with  m$_{ax}$= 8.5 meV 
}
\end{figure}
\begin{figure}
\epsfxsize=8cm
\epsfbox{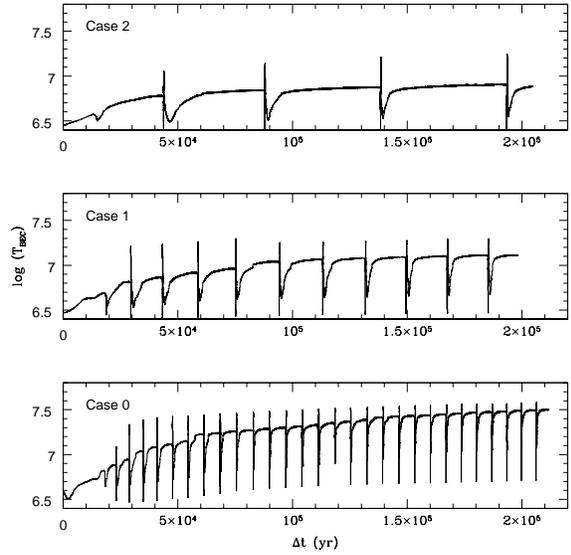}
\caption{Evolution with time of the temperature at the botton of 
the convective envelope at the TP-AGB phase, 
  for the 5M$_\odot$ star.
Lower panel: m$_{ax}$=0.0 (Case 0);
 Middle panel: m$_{ax}$=8.5 meV (Case 1); Upper panel:
  m$_{ax}$=20 meV (Case 2).}
\end{figure}

More He is accumulated by the H-burning during the extended interpulse period, 
so that more nuclear energy must be produced in the pulse in order to expand the stored matter.
A stronger thermal pulse favours the following dredge-up.     
For this reason the third  dredge-up starts after the first TP in  the two sequences including axions,
while in Case 0 it starts after the $4^{th}$ TP. 
The mass of He and C  material mixed in the envelope is nearly twice as much 
  in Case 1 and
 $\sim$8 times larger in Case 2, both with respect to Case 0.

With a resolution of just 1M$_\odot$,
the minimum mass for which carbon ignition occurs before the TP-AGB phase, i.e. M$_{up}$,
is the same in all three cases, namely $\sim$8 $M_\odot$.
However the C ignition occurs farther from the centre in models with axions.
 For instance, in Case 1 we found an off centre C ignition at 0.93M$_\odot$ instead
 of the 0.40M$_\odot$ in  the standard case.
Note that the mass of the CO core is smaller when axions are included, namely 
1.148 M$_\odot$ instead of 1.259 M$_\odot$.

\setcounter{table}{1}
\begin{table*}
\begin{minipage}{160mm}
\caption{Properties of the 5M$_\odot$ model during the TP-AGB phase}
\begin{tabular}{@{}ccccc}
    & & & &  \\
M$_H$ &
$\Delta$t$_{ip}$ &
$\Delta$M$_{H}$ &
$\Delta$M$_{D-up}$ &
$\lambda$ \\
    & & & &  \\
Case 2: 20 meV   &   &  &  &  \\
    & & & &  \\
0.661   & ... & ... & 2.50E-03 & ... \\
0.666   & 4.49E+04 & 7.80E-03 & 6.50E-03 & 0.83\\
0.670   & 4.96E+04 & 1.03E-02 & 8.80E-03 & 0.85\\
0.673   & 5.68E+04 & 1.17E-02 & 1.02E-02 & 0.87\\
    & & & &  \\
Case 1: 8.5 meV   &   &  &  &  \\
    & & & &  \\
0.748   & ... & ... & ... & ... \\
0.750   & 1.37E+04 & 2.40E-03 & 0.50E-03 & 0.21\\
0.753   & 1.54E+04 & 3.10E-03 & 1.30E-03 & 0.42\\
0.755   & 1.66E+04 & 3.60E-03 & 1.70E-03 & 0.47\\
0.757   & 1.89E+04 & 3.90E-03 & 2.00E-03 & 0.51\\
0.760   & 1.88E+04 & 4.20E-03 & 1.90E-03 & 0.45\\
0.762   & 1.85E+04 & 4.20E-03 & 1.80E-03 & 0.43\\
0.765   & 1.82E+04 & 3.90E-03 & 1.70E-03 & 0.44\\
0.767   & 1.78E+04 & 4.10E-03 & 2.00E-03 & 0.49\\
0.769   & 1.77E+04 & 4.20E-03 & 1.80E-03 & 0.43\\
    & & & &  \\
Case 0: 0  meV   &   &  &  &  \\
    & & & &  \\
0.860   & ... & ... & ... & ... \\
0.861   & 4.84E+03 & 1.03E-03 & ... & ... \\
0.862   & 5.70E+03 & 1.24E-03 & ... & ... \\
0.864   & 6.10E+03 & 1.30E-03 & 1.33E-04 & 0.10 \\
0.865   & 6.34E+03 & 1.57E-03 & 3.26E-04 & 0.21\\
0.867   & 6.60E+03 & 1.72E-03 & 4.80E-04 & 0.28\\
0.868   & 6.80E+03 & 1.85E-03 & 5.56E-04 & 0.30\\
0.869   & 6.93E+03 & 1.93E-03 & 6.61E-04 & 0.34\\
0.871   & 7.05E+03 & 2.02E-03 & 7.99E-04 & 0.40\\
0.872   & 7.12E+03 & 2.10E-03 & 8.60E-04 & 0.41\\
0.873   & 7.16E+03 & 2.14E-03 & 8.75E-04 & 0.41\\
0.875   & 7.16E+03 & 2.17E-03 & 8.85E-04 & 0.41\\
0.876   & 7.14E+03 & 2.20E-03 & 9.94E-04 & 0.45\\
0.877   & 7.13E+03 & 2.23E-03 & 9.06E-04 & 0.41\\
0.878   & 7.07E+03 & 2.22E-03 & 1.00E-03 & 0.45\\
0.881   & 7.00E+03 & 2.29E-03 & 1.07E-03 & 0.47\\
0.882   & 6.99E+03 & 2.28E-03 & 1.17E-03 & 0.51\\
0.883   & 6.97E+03 & 2.30E-03 & 1.22E-03 & 0.53\\
0.884   & 6.95E+03 & 2.32E-03 & 1.24E-03 & 0.53\\
0.885   & 6.90E+03 & 2.32E-03 & 1.21E-03 & 0.52\\
0.886   & 6.84E+03 & 2.31E-03 & 1.25E-03 & 0.54\\
0.887   & 6.80E+03 & 2.32E-03 & 1.26E-03 & 0.54\\
0.888   & 6.73E+03 & 2.31E-03 & 1.28E-03 & 0.55\\
0.889   & 6.70E+03 & 2.32E-03 & 1.29E-03 & 0.56\\
0.891   & 6.66E+03 & 2.32E-03 & 1.23E-03 & 0.53\\
0.892   & 6.56E+03 & 2.29E-03 & 1.17E-03 & 0.51\\
0.893   & 6.45E+03 & 2.25E-03 & 1.19E-03 & 0.53\\
0.894   & 6.41E+03 & 2.25E-03 & 1.28E-03 & 0.57\\
    & & & &  \\
\end{tabular}
\end{minipage}
\end{table*}

Finally, it is interesting to note  that the inclusion of axions modifies
the initial/final mass relation (see e.g. Weidemann 1987). During the thermal 
 pulses, the mass of the CO core increases as a result of the accretion of 
 C freshly synthesized by the He flashes. However, the observed luminosity 
 of AGB stars (Weidemann 1987; Wood et al. 1992) and the observed mass-loss 
 rate at this phase, indicate that the occurrence of too many pulses is 
 prevented by the removal of the envelope (Bedijn 1987, Baud and Habing 1983, 
 van der Veen, Habing and Geballe 1989). The mass of the CO white dwarf is 
 essentially determined by the value reached at the end of the E-AGB phase 
 (see Vassiliadis and Wood 1993; Bl\"ocker 1995). 
   In figure 8 we compare the CO core mass (a  guess for the WD mass) with the 
  initial mass (progenitor mass), for Case 0 and Case 1. Two more masses have 
 been included, 4 and 6 M$_\odot$.

   The relative abundance of
white dwarfs with  masses $m_0$ and $m_1$ is given by
$$
r(m_1,m_0) = \frac{\delta n(m_1)}{\delta n(m_0)}=
\Big[\frac{M(m_0)}{M(m_1)}\Big]^{\alpha}\,\frac{(dM/dm)_{m_0}}{(dM/dm)_{m_1}}
$$
\noindent
where we have assumed that the star formation rate per unit volume in the
solar neighbourhood is constant and that the initial mass function
follows Salpeter's law with $\alpha = 2.35$ and M(m) is the relationship
 between  the mass of the parent star and that  of the white dwarf.
Note that if it were possible to separate the influence of the mass loss
and the influence of axions during the AGB phase, it would be possible to
obtain very strong constraints to the properties of these particles.

\begin{figure}
\epsfxsize=8cm
\epsfbox{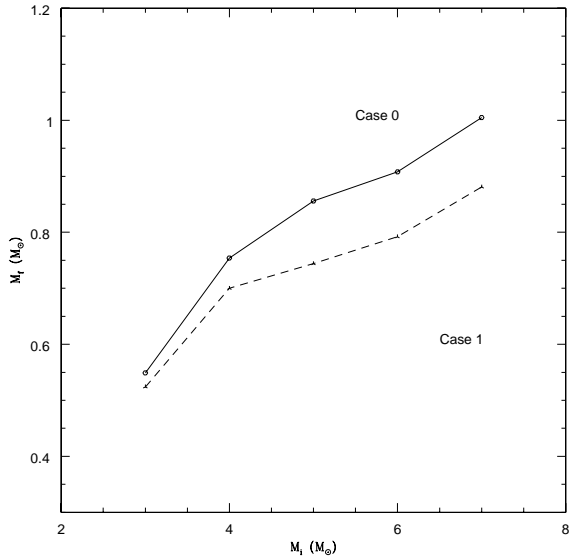}
\caption{
Final and Initial masses;  solid line=
Case 0 (m$_{ax}$=0.0)
  and dashed line= Case 1 (m$_{ax}$=8.5 meV).
}
\end{figure}

\section{Conclusions}

We have studied the effects of including axions (DFSZ model) in the
evolution of intermediate and low mass stars ($0.8 \leq M/M_\odot \leq 9$). 
Among the various evolutionary phases affected by  axion emission, the AGB
seems to be the most promising in terms of obtaining  stringent constraints to the theory
of particle physics. These effects are: 

1.- A severe reduction in the size of the final CO core, as compared with the
values obtained from standard evolutions, implies some interesting modifications of 
observable quantities, such as the luminosity of AGB stars and their residual mass.
 Concerning the AGB luminosity function in particular, the axion emission would imply 
a deficit of bright AGB stars. The initial/final mass relation is also substantially modified and  
  a deficit of massive white dwarfs is expected.

2.- As a consequence of the axion cooling, the convective instabilities that characterize the
double nuclear shell burning are more extended and the chemical composition
of the surface could be significantly modified. A stronger
$3^{rd}$ dredge-up implies that 
the C star stage could be more easily (and perhaps rapidly) obtained. In addition the
surface abundance of s-elements, which are the products of the neutron capture
nucleosynthesis occurring during the interpulse (through the
$^{13}C(\alpha,n)$ neutron source) and during the thermal pulse (through the
$^{22}N(\alpha,n)$), are other important "observable" quantities to be used to constrain the 
properties of the axions as well as those of any other weak interactive particle proposed in
non-standard theories of particle physics.

It is a great pleasure to thank Alessandro Chieffi and Marco Limongi
for helpful discussion.
This work has been supported in part by the DGICYT grants PB96-1428 and 
ESP98-1348
, by the Italian Minister of University,  Research and Technology (Stellar
 Evolution project),
by the Junta de Andalucia  FQM-108 project and by the CIRIT(GRC/PIC).

\end{document}